\documentclass[aps,twocolumn,preprintnumbers,superscriptaddress,prl,10pt]{revtex4-2}
\usepackage{bbm}
\usepackage{bm}
\usepackage{mathrsfs }
\usepackage{amsmath}
\usepackage{amssymb}
\usepackage{empheq}
\usepackage{graphicx}
\usepackage{mathrsfs}
\usepackage{amsfonts}
\usepackage{amsthm}
\usepackage{color}
\usepackage{multirow}
\usepackage{bigints}
\usepackage{txfonts}
\usepackage{float}
\usepackage{hyperref}
\hypersetup{
     unicode=false,              
     pdftoolbar=true,           
     pdfmenubar=true,         
     pdffitwindow=false,      
     pdfstartview={FitH},    
     pdftitle={My title},       
     pdfauthor={Author},     
     pdfsubject={Subject},   
     pdfcreator={Creator},   
     pdfproducer={Producer}, 
     pdfkeywords={keyword1} {key2} {key3}, 
     pdfnewwindow=true, 
     colorlinks=false,       
     linkcolor=red,           
     citecolor=green,        
     filecolor=magenta,    
     urlcolor=cyan           
}

\makeatletter \tolerance = 10000 \tolerance = 10000

\makeatother
\begin{document}
\title{Curvature-Induced Geometric Universality in Non-Hermitian Anderson Transitions}

\author{Chen Wang}
\email[Contact author: ]{physcwang@tju.edu.cn}
\affiliation{Center for Joint Quantum Studies and Tianjin Key Laboratory of Low Dimensional Materials Physics and Preparing Technology, Department of Physics, School of Science, Tianjin University, Tianjin 300350, China}
\author{Run-Qiu Yang}
\email[Contact author: ]{aqiu@tju.edu.cn}
\affiliation{Center for Joint Quantum Studies and Tianjin Key Laboratory of Low Dimensional Materials Physics and Preparing Technology, Department of Physics, School of Science, Tianjin University, Tianjin 300350, China}
\author{X. R. Wang}
\email[Contact author: ]{phxwan@cuhk.edu.cn}
\affiliation{School of Science and Engineering, Chinese University of Hong Kong (Shenzhen), Shenzhen 518172, China}
\author{Hechen Ren}
\email[Contact author: ]{ren@tju.edu.cn}
\affiliation{Center for Joint Quantum Studies and Tianjin Key Laboratory of Low Dimensional Materials Physics and Preparing Technology, Department of Physics, School of Science, Tianjin University, Tianjin 300350, China}

\date{\today}

\begin{abstract}
In Euclidean space, universality classes of Anderson transitions are primarily determined by symmetry and spatial dimensionality. Here, we present evidence for a geometry-controlled universality class of non-Hermitian Anderson transitions on hyperbolic-like lattices. In this setting, critical behavior is influenced by the large-scale hyperbolic geometry, characterized by negative curvature, exponential volume growth, and a non-Euclidean notion of spatial scaling. Finite-size scaling of participation ratios across several distinct \( \{p,q\} \) tilings reveals one-parameter scaling collapses with a common critical exponent \( \nu\simeq1 \) within numerical accuracy. A complementary phenomenological coarse-grained Landau-Ginzburg analysis shows how exponential correlation-volume growth suppresses critical fluctuations, offering a rationale for the observed mean-field-like scaling. Our results suggest that spatial curvature can act as an additional organizing principle for Anderson-transition universality beyond the conventional dimensionality- and symmetry-based classification.
\end{abstract}

\maketitle

\emph{Introduction.}$-$Anderson transition, a disorder-driven transition between extended and localized phases, is one of the central paradigms in condensed-matter physics~\cite{pwAnderson_pr_1958}. In Euclidean space, its critical behaviors are controlled not only by the system's symmetry (orthogonal/unitary/symplectic), but also sensitively by geometric properties, most prominently the spatial dimension \( d \)~\cite{eAbrahams_prl_1979,bKramer_rpp_1993,bHuckestein_rmp_1995,fEvers_rmp_2008}. Accordingly, the correlation-length critical exponent \(\nu\) takes distinct values in two, three, and higher dimensions, thereby underpinning the conventional universality classification of Anderson transitions~\cite{dfriedan_prl_1980,shimami_prp_1980,ammpruisken_prl_1988,yasada_prl_2002,jLi_prl_2009,cwang_prl_2015,czChen_prl_2015,LUjfalusi_prb_2015,bFu_prl_2017,twang_prb_2021,caLi_prl_2026}.
\par

By contrast, much less is known about Anderson transitions in non-Euclidean spaces, particularly on negatively curved hyperbolic lattices~\cite{aChen_prl_2024,Li_cp_2024,bCurtis_prl_2025,LShou_prl_2025,ySun_ps_2026,aaltland_arxiv_2026}. Unlike the polynomial volume growth \(N(L)\sim L^d\) of Euclidean lattices, hyperbolic lattices exhibit exponential growth, \(N(L)\sim \eta^L\), where \(N\) is the total number of sites within graph distance \(L\) from a chosen origin and \(\eta>1\) is a geometry-dependent factor~\cite{anderson_2005_hyperbolic,ratcliffe_2006_foundations,beardon_2012_geometry}. This non-Euclidean connectivity profoundly modifies spectral and transport properties, motivating recent studies of hyperbolic band structures~\cite{jMaciejko_sa_2021,jMaciejko_pnas_2022,nCheng_prl_2022,mLenggenhager_prl_2023}, flat bands~\cite{mUrwyler_prl_2022,bvsfiek_prb_2022,yhao_prb_2024,dguan_prb_2025}, topological phases~\cite{syu_prl_2020,pBienias_prl_2022,aStegmaier_prl_2022,ytao_prb_2023,tTummuru_prl_2024,kGyunghun_prl_2026,zliu_prb_2026,jmChen_nc_2026}, and many-body physics~\cite{xzhu_jpc_2021,mlenggenhager_prl_2025,ahe_prb_2025,cleong_prb_2026}.
\par

\begin{figure}[htbp]
\includegraphics[width=0.38\textwidth]{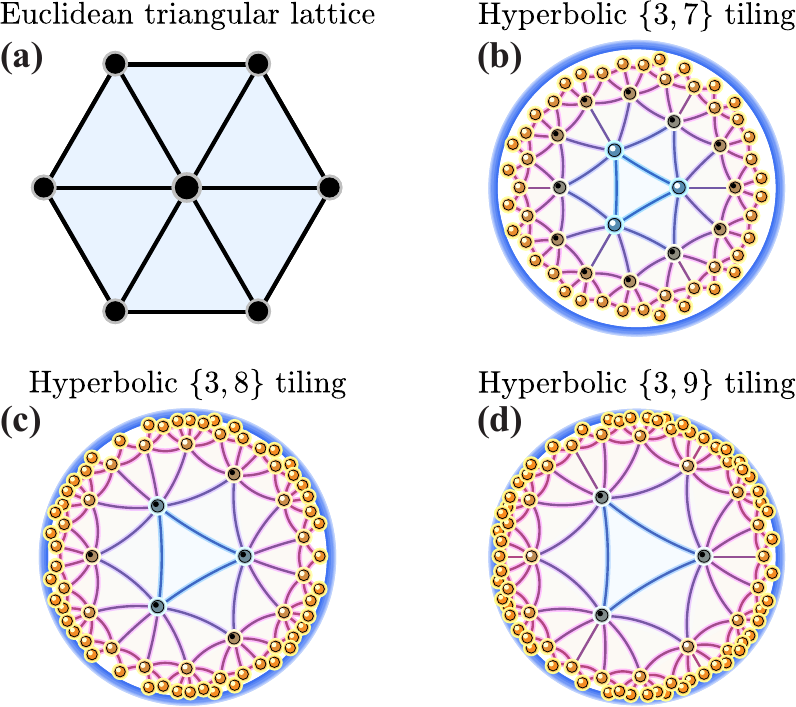}\centering
\caption{Schematic comparison of a Euclidean triangular lattice and three representative \( \{p,q\} \) hyperbolic tilings, where each polygon has \( p \) sides and \( q \) polygons meet at each vertex: (a) the triangular lattice with \( \{p,q\}=\{3,6\} \); (b–d) hyperbolic tilings with \( \{p,q\}=\{3,7\} \), \( \{3,8\} \), and \( \{3,9\} \), respectively. While the Euclidean lattice is flat, the hyperbolic tilings possess negative curvature and exhibit exponential volume growth, \( N(L)\sim \eta(p,q)^L \). Finite-size scaling analyses indicate that the non-Hermitian Anderson transitions on these hyperbolic-like lattices share a common critical exponent \( \nu\simeq 1 \).}
\label{fig0}
\end{figure}

Although Hermitian Anderson transitions have been studied on selected hyperbolic lattices~\cite{aChen_prl_2024,Li_cp_2024,aaltland_arxiv_2026}, how geometric parameters, such as those defining the \(\{p,q\}\) tilings shown schematically in Fig.~\ref{fig0}, affect their universality remains unclear. The situation is even less explored for non-Hermitian Anderson transitions, and the orthodox universality framework cannot readily extend to curved systems that lack a well-defined finite spatial dimensionality. This challenge is particular relevant in view of the distinctive delocalization phenomena enabled by non-Hermiticity, including extended states in one dimension~\cite{nHatano_prl_1996,dLeykam_prl_2017,yap_prl_2018,vmMartinez_prb_2018,hSun_prl_2018,kKawabata_prl_2021,wWang_prl_2025,cWang_prl_2025,sLonghi_prl_2025,bLi_prl_2025,tyang_prl_2025,cWang_prb_2025,wtXue_prl_2026,sqLi_prb_2026}. These considerations thus raise a fundamental open question: can negative curvature and non-Hermiticity generate new universality classes of Anderson transitions?
\par

In this article, we numerically investigate non-Hermitian Anderson transitions on hyperbolic-like lattices. Using finite graphs that faithfully approximate regular \(\{p,q\}\) tilings, we perform exact diagonalization and finite-size scaling analyses based on the participation ratio \(p_2\). Across all considered tilings, we identify clear Anderson transitions, characterized by robust one-parameter scaling collapses and a common critical exponent \(\nu\simeq 1\) within numerical uncertainty. These results indicate that critical behaviors of non-Hermitian Anderson transitions on hyperbolic-like lattices are insensitive to microscopic tiling details and other nonuniversal features, as long as the underlying geometry remains negatively curved.
\par

We refer to this phenomenon as \emph{geometric universality}. A coarse-grained Landau-Ginzburg analysis suggests that this universality arises from hyperbolic space's exponential volume growth, which strongly suppresses critical fluctuations. The resulting critical behavior is therefore mean-field-like, with \(\nu\simeq1 \) remaining insensitive to the specific hyperbolic tiling. Our results thus establish geometry as an independent organizing principle for Anderson-transition universality in curved non-Hermitian systems.
\par

\emph{Hyperbolic-like graphs.}$-$To construct the hyperbolic-like lattices, we begin from a regular hyperbolic tiling \( \{p,q\} \), in which each face is a regular hyperbolic \( p \)-gon and exactly \( q \) polygons meet at every vertex, satisfying the condition~\cite{anderson_2005_hyperbolic}
\begin{equation}
    \begin{gathered}
        1/p+1/q<1/2
    \end{gathered}\label{eq_1}
\end{equation}
with \( p,q\geq 3 \) being integers. The tiling is embedded in the Poincar\'{e} disk \( \mathbb{D} \) so that vertices map to points \( (x_i,y_i)\in\mathbb{D} \) and edges to geodesic arcs. For numerical calculations, we restrict the infinite tiling to a finite radial patch. Choosing a central face, we include all faces whose distance from it in the dual graph is at most \(L\)~\cite{mSchrauth_sp_2024}. The vertices incident on these faces define a finite set \( \mathcal{V}_L=\{ 1,2,\cdots, N_L\} \). Two vertices are connected by an undirected edge \( \mathcal{E}_L \) if they are nearest neighbors, which yields an open-boundary graph \( G_L=(\mathcal{V}_L,\mathcal{E}_L) \). 
\par

To approximate homogeneous hyperbolic lattices, we add controlled extra edges between under-coordinated boundary sites, chosen to drive vertex degrees toward \(q\) and to close short cycles of length \(\simeq p\). Such hyperbolic-like graphs, denoted as \(\tilde{G}_L=(\tilde{\mathcal{V}}_L,\tilde{\mathcal{E}}_L)\), have a degree distribution sharply peaked near \(q\), with averaged curvature and topology consistent with an ideal \(\{p,q\}\) tiling~\cite{supp}. In Table~\ref{table_1}, we list the effective tiling parameters \( \tilde{p} \) and \( \tilde{q} \) for all lattices studied. Although this graph-level gluing does not constitute an exact compact quotient of the hyperbolic plane, the close agreement between \( \{\tilde p,\tilde q\} \) and the target \( \{p,q\} \) values indicates that the local coordination and face structure are well preserved. 
\par

\begin{table}
\caption{This table summarizes the finite-size scaling analysis for various hyperbolic-like tilings. The first and second columns list the target and effective tilings, denoted by \( \{p,q\} \) and \( \{\tilde{p},\tilde{q}\} \), respectively. The third, fourth, fifth, and sixth columns report the critical disorder \( W_c \), the effective fractal dimension \( D \), the critical exponent \( \nu \), and the goodness-of-fit \( Q \). Other parameters are given in Supplemental Materials~\cite{supp}.} 
\begin{ruledtabular}
\begin{tabular}{cccccc}
\( \{p,q\} \) & \( \{ \tilde{p},\tilde{q} \} \) & \( W_c \) & \( D \) & \( \nu \) & Q \\
\hline
\( \{3,7\} \)  & \( \{3.12,7\} \)  & \( 5.8\pm0.2 \)   & \( 1.2\pm0.2 \)   & \( 1.00\pm0.04 \)  & 0.6 \\
\( \{3,8\} \)  & \( \{3.13,8\} \)  & \( 6.2\pm0.3 \)   & \( 1.3\pm0.2 \)   & \( 1.00\pm0.08 \)  & 0.3 \\
\( \{3,9\} \)  & \( \{3.16,9\} \)  & \( 6.6\pm0.3 \)   & \( 1.3\pm0.2 \)   & \( 1.0\pm0.1 \)    & 0.5 \\
\( \{3,10\} \) & \( \{3.12,10\} \) & \( 6.8\pm0.4 \)   & \( 1.3\pm0.1 \)   & \( 0.95\pm0.09 \)  & 0.1 \\
\( \{3,11\} \) & \( \{3.03,11\} \) & \( 7.3\pm0.2 \)   & \( 1.1\pm0.1 \)   & \( 1.0\pm0.2 \)    & 0.4 \\
\( \{3,12\} \) & \( \{3.01,12\} \) & \( 7.6\pm0.2 \)   & \( 1.1\pm0.2 \)   & \( 1.05\pm0.09 \)  & 0.05 \\
\( \{7,3\} \)  & \( \{7.01,3\}  \) & \( 3.25\pm0.06 \) & \( 1.1\pm0.1 \)   & \( 0.95\pm0.06 \)  & 0.6 \\
\( \{8,3\} \)  & \( \{8.05,3\}  \) & \( 3.3\pm0.4 \)   & \( 1.3\pm0.1 \)   & \( 1.02\pm0.07 \)  & 0.1 \\
\end{tabular}
\end{ruledtabular}\label{table_1}
\end{table}

\emph{Tight-binding models.}$-$Then, we consider a single-orbital tight-binding model with nearest-neighbor hopping on the above glued hyperbolic-like graphs. The Hilbert space is spanned by localized orbitals \( |i\rangle \), where \( i\in \tilde{\mathcal{V}}_L \), and the lattice Hamiltonian reads
\begin{equation}
    \hat{H}=\sum_{i}(\epsilon^{\text{R}}_i+i\epsilon^{\text{I}}_i) c^\dagger_i c_i+ \sum_{\langle ij \rangle}(tc^\dagger_i c_j+\mathrm{h.c.}),
    \label{eq_2}
\end{equation}
where \( \langle ij\rangle \) runs over all the nearest-neighbor pairs in \( \tilde{\mathcal{E}}_L \). We set \( t=1 \) as the energy unit and introduce a complex Anderson-type disorder by setting \( \epsilon^{\text{R}}_i \) and \( \epsilon^{\text{I}}_i \) to be distributed independently and uniformly within \( [-W/2, W/2] \)~\cite{bKramer_rpp_1993}. Thus, \( W \) measures the disorder strength. 
\par

\emph{Finite-size scaling analysis.}$-$To identify Anderson transitions, we use the disorder-averaged participation ratio, \( p_2=(\sum_i |\psi_i(E)|^4)^{-1} \) at a fixed energy \(E=0\), to perform the finite-size scaling analysis. Here, \(\psi_i(E)\) is the amplitude on the \(i\)th site of the normalized right eigenstate of model~\eqref{eq_2}. Near the critical disorder \(W_c\), we assume the scaling function~\cite{jhpixley_prl_2015,cWang_prb_2025}
\begin{equation}
    p_2(\tilde{L},W)=g(\tilde{L}) f(\tilde{L}/\xi)+\tilde{L}^{-\tilde{y}}\tilde{f}(\tilde{L}/\xi).
    \label{eq_3}
\end{equation}
Here, \( g(\tilde{L}) \) represents the scaling of the wave function at \(W_c\). \(f(x)\) and \( \tilde{f}(x) \) stand for the relevant and irrelevant scaling functions, respectively. The correlation length \( \xi \) is assumed to diverge as \(\xi\sim |W-W_c|^{-\nu}\), where \(\nu\) is the critical exponent characterizing the universality class, and, for hyperbolic-like lattices, the effective linear length is commonly taken as \(\tilde L=\ln N\)~\cite{aChen_prl_2024,Li_cp_2024}. This scaling ansatz describes our data well; moreover, an alternative two-sided scaling analysis gives consistent estimates of \(\nu\)~\cite{iGarca_prb_2022}. The positive constant \( \tilde{y} \) stands for the irrelevant exponent~\cite{cWang_prb_2025}.
\par

After fitting our data to Eq.~\eqref{eq_3}, we extract \(W_c\), \(\nu\), \(D\), and other fitting parameters and obtain the reduced participation ratio defined as \( Y_{\tilde L}=(p_2-\tilde{L}^{-\tilde{y}}\tilde{f}(\tilde{L}/\xi))/g(\tilde L) \) such that \(Y_{\tilde L}\) depends only on \(\tilde L/\xi\)~\cite{cWang_prb_2025}. Therefore, curves of \(Y_{\tilde L}\) for different system sizes cross at the critical disorder \(W_c\) and collapse onto a one-parameter scaling function near the transition with a suitable scaling variable \( x=\tilde{L}^{1/\nu}(W-W_c) \). 
\par

An important issue is the appropriate choice of \(g(\tilde L)\). For hyperbolic-like lattices, two natural possibilities arise. The first is a length-based scaling, \(g(\tilde L)\sim \tilde L^D\), in which the critical participation ratio is governed by the effective linear length~\cite{aChen_prl_2024}. The second is a volume-based scaling, \(g(\tilde L)\sim N^{D'}\), in which the scaling is governed by the total number of lattice sites \( N \) characterized by a different dimension \( D' \). Through analyses of data by the two different forms, we find that the length-based scaling yields a satisfactory goodness-of-fit \(Q\) in the \(\chi^2\) analysis (\( Q>10^{-3} \))~\cite{nr_press,supp}. Therefore, the critical behavior of our data is more naturally controlled by \( \tilde{L} \) rather than by \(N\), i.e., \(g(\tilde L)\sim \tilde L^D\).
\par

\begin{figure}[htbp]
\includegraphics[width=0.48\textwidth]{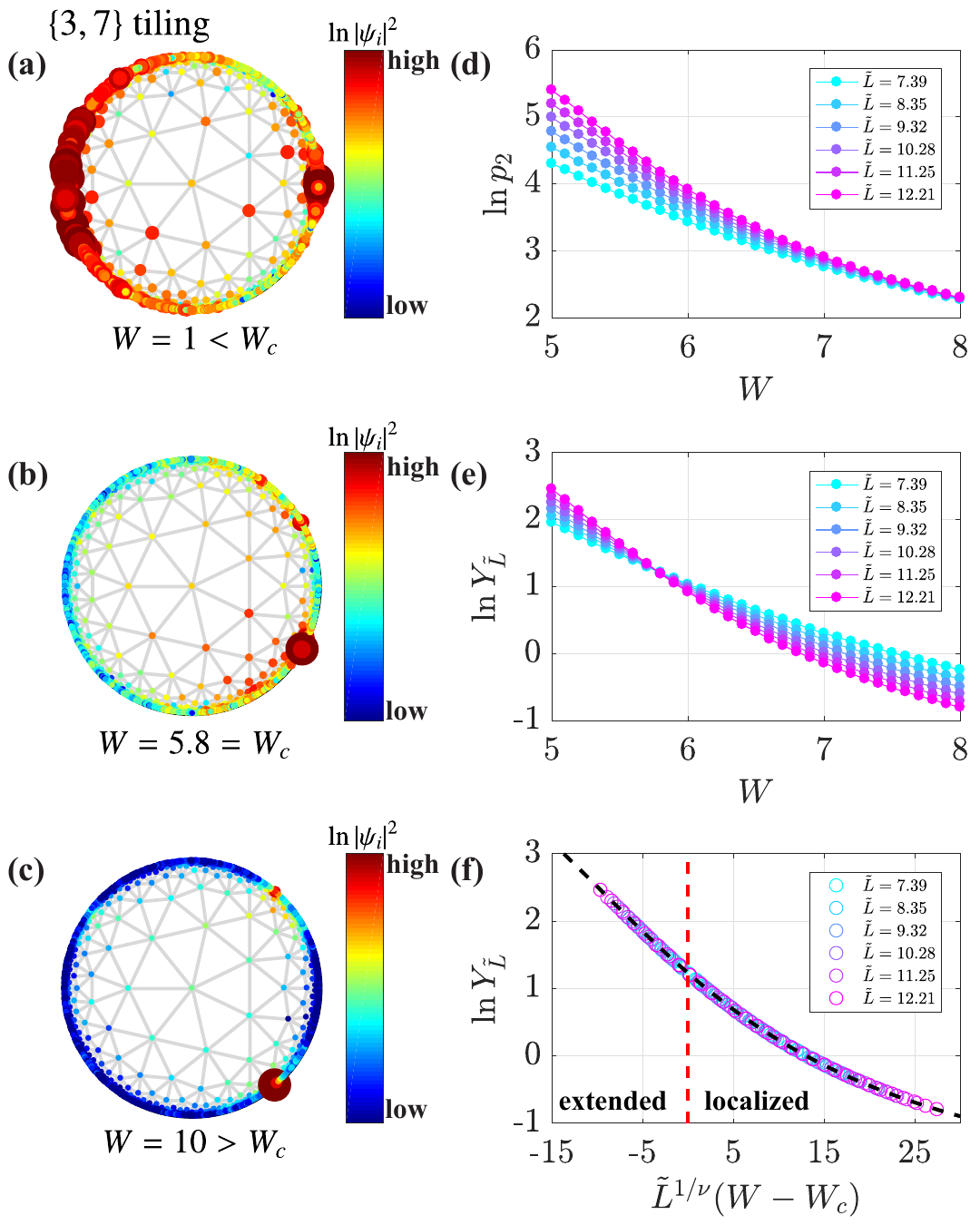}\centering
\caption{Plots of the wave‑function amplitude distributions on the \(\{3,7\}\) lattice and \( E=0 \) for (a) the extended phase at \(W=1<W_c\), (b) the critical point at \(W=5.8=W_c\), and (c) the localized phase at \(W=10>W_c\). The color scale encodes \(\ln|\psi_i|^2\). One ensemble is used here. (d) \(\ln p_2\) as a function of \(W\) for various \( \tilde{L} \) of the \(\{3,7\}\) lattice at energy \(E=0\). (e) \( \ln Y_{\tilde{L}} \) versus \(W\) for the same data as in (d). (f) Scaling collapse of \(\ln Y_{\tilde{L}} \) as a function of the scaling variable \( x=\tilde{L}^{1/\nu}(W-W_c) \). The black dashed line denotes the scaling function \( f(x) \) obtained from the \(\chi^2\) fit, while the red solid line marks the phase boundary between the extended and localized phases.}
\label{fig1}
\end{figure}

\emph{Anderson transitions.}$-$Anderson transitions can happen on hyperbolic-like lattices. As one typical example, Figs.~\ref{fig1}(a), (b), and (c) display the natural logarithm of the wave‑function intensity \( \ln{[|\psi_i|^2]} \) for a single disorder realization at three values of \( W=1 \), 5.8, and 10, respectively, for the \( \{3,7\} \) tiling. For \( W=1<W_c \) and \( W=10>W_c \), the state spreads over the entire sample and concentrates in a few isolated sites, respectively. At the critical disorder \( W_c \) (determined below), the wave function exhibits a highly inhomogeneous, self‑similar pattern characteristic of a multifractal critical state~\cite{fEvers_rmp_2008}.
\par

The corresponding finite-size scaling analysis is shown in Figs.~\ref{fig1}(d), (e), and (f). Figure~\ref{fig1}(d) depicts \( \ln{p_2} \) as a function of \( W \) for \( \tilde{L} \) ranging from \( 7.39 \) to \( 12.21 \). Figure~\ref{fig1}(e) shows the corresponding \( \ln{Y_{\tilde{L}}} \), from which one can clearly see a crossing point at \( W_c=5.8\pm0.2 \), signaling an Anderson transition. Near the critical point, we plot the scaling function in Fig.~\ref{fig1}(f), i.e., \( \ln{Y_{\tilde{L}}}=\ln{[f(x=\tilde{L}^{1/\nu}(W-W_c))]}\) with \( \nu=1.00\pm0.04 \). The smooth scaling function in Fig.~\ref{fig1}(f) strongly supports our one-parameter scaling function~\eqref{eq_3}. 
\par

\begin{figure}[htbp]
\includegraphics[width=0.45\textwidth]{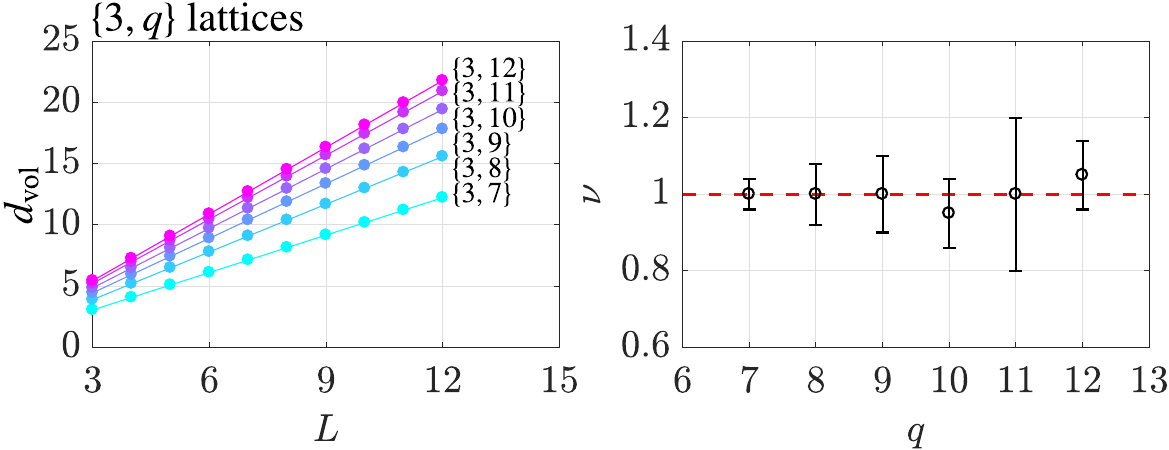}\centering
\caption{(a) Volume-growth dimension \( d_{\rm vol} \) of the \(\{3,q\}\) hyperbolic-like lattices as a function of graph distance \( L \) from the center. The slopes of these curves determine the volume-growth factors \( \eta(3,q) \). (b) Critical exponent \( \nu \) for the \(\{3,q\}\) tilings, which is consistent with \( \nu = 1 \) (red dashed line) within numerical uncertainty.}
\label{fig2}
\end{figure}

\emph{Geometric universality.}$-$Next, we provide evidence for geometric universality. In general, one can define the volume-growth dimension as \( d_{\rm vol}=\mathrm{d}\ln N/\mathrm{d}\ln L \). In Euclidean space, \( N\sim L^d \) such that the volume-growth dimension \( d_{\text{vol}} \) equals the spatial dimension \( d \). However, for hyperbolic-like lattices, the volume-growth dimension becomes size-dependent:
\begin{equation}
    \begin{gathered}
        d_{\rm vol}=L\ln{[\eta(p,q)]},
    \end{gathered}\label{eq_4}
\end{equation}
since \( N\sim\eta(p,q)^{L} \). Therefore, hyperbolic-like lattices are effectively infinite-dimensional for large-scale graph distance \( L\to\infty \). Moreover, different \(\{p,q\}\) tilings have different growth factors \(\eta(p,q)\), leading to distinct volume-growth rates as a function of graph distance \(L\), as shown in Fig.~\ref{fig2}(a).
\par

As the Euclidean universality classes are strongly affected by changes in the spatial dimension \( d \), or equivalently by the volume-growth dimension \(d_{\mathrm{vol}}\)~\cite{fEvers_rmp_2008}, it is natural to ask whether the critical behavior on hyperbolic-like lattices is similarly sensitive to the volume-growth factor \( \eta(p,q) \). To address this question, we perform additional finite-size scaling analyses on \(\{3,q\}\) tilings with \(q=8,9,\ldots,12\). Remarkably, our results lead to a different behavior: within numerical uncertainty, all data are consistent with \( \nu\simeq 1 \), despite the substantial variation of the growth factor \(\eta(3,q)\); see Fig.~\ref{fig2}(b). We further study some dual tilings like \(\{7,3\}\) and \(\{8,3\}\), and our simulations consistently yield \(\nu = 1\); see Table~\ref{table_1}. 
\par

The common critical exponent \(\nu\) provides strong evidence for a shared critical behavior of non-Hermitian Anderson transitions on hyperbolic-like lattices. Although the local connectivity, volume-growth factor \(\eta(p,q)\), and critical disorder strength \(W_c\) vary among different tilings, \(\nu\) remains unchanged within numerical uncertainty. Consistently, the fractal dimension \(D\), which is generally expected to be universal within a given Euclidean universality class~\cite{fEvers_rmp_2008}, remains nearly identical across the considered tilings, with \(D\in[1.1,1.3]\). We refer to this robust insensitivity to microscopic tiling details as geometric universality, since the universal critical properties are determined not by the specific volume-growth dimension \(d_{\mathrm{vol}}\), but by the common large-scale hyperbolic geometry.
\par

\emph{Robustness check.}$-$To assess the robustness of the observed geometric universality, we perform three independent consistency checks on the \(\{3,7\}\) tiling: (i) varying the onsite disorder distribution, (ii) using the biorthogonal definition of \(p_2\), and (iii) randomizing the lattice connectivity while preserving negative curvature. In all cases, the fitted exponent remains consistent with \(\nu=1\) within numerical uncertainty, indicating that the observed geometric universality is not tied to a specific participation-ratio definition, disorder ensemble, or microscopic connectivity pattern. Detailed evidence is provided in Supplemental Material~\cite{supp}.
\par

\emph{Origin of geometric universality.}$-$We provide a field-theoretic rationale for the appearance of geometric universality from a scalar-field Ginzburg analysis. We begin from the metric of hyperbolic space in the Poincar\'{e} disk:
\begin{equation}
    \begin{gathered}
        \mathrm{d}s^2=4(\mathrm{d}r^2+r^2\mathrm{d}\theta^2)/(1-r^2)^2,
    \end{gathered}\label{eq_5}
\end{equation}
where \( (r,\theta) \) are polar coordinates in the Poincar\'{e} disk and \( r<1 \). The Gaussian curvature \( K \) can be computed directly from the metric components \( g_{rr}=4/(1-r^2)^2 \), \( g_{\theta\theta}=4r^2/(1-r^2)^2 \), and \( g_{r\theta}=g_{\theta r}=0 \) via the standard formula:~\cite{supp}
\begin{equation}
    \begin{gathered}
        K=R_{r\theta r\theta}/(g_{rr}g_{\theta\theta})=-1.
    \end{gathered}\label{eq_6}
\end{equation}
Here, \( R_{r\theta r\theta} \) is the relevant component of the Riemann curvature tensor, constructed from the Christoffel symbols of the metric. The constant negative Gaussian curvature is the defining geometric feature of the continuum hyperbolic plane.
\par

We consider a complex massive scalar field \( \psi \) in hyperbolic space, whose dynamics is governed by the Klein-Gordon-type equation \( \left( -\nabla^2_g +m^2(x)\right)\psi=0 \) with \( \nabla^2_g \) being the Laplace-Beltrami operator~\cite{buser_2010_geometry}. The corresponding quadratic action for the scalar field reads
\begin{equation}
    \begin{gathered}
        S[\psi]=\dfrac{1}{2}\int_{\mathbb{D}}\mathrm{d}^2x \sqrt{g}\psi^\ast(-\nabla^2_g+m^2(x))\psi, 
    \end{gathered}\label{eq_7}
\end{equation}
where \( g \) denotes the determinant of the metric tensor. The randomness is encoded in the mass term by setting \( m^2(x)=m^2_0+\epsilon(x) \) with \(m_0\) being a constant mass and \( \epsilon(x) \) being a zero-mean complex random field. In our model, the disorder correlations are \( \langle \epsilon(x)\epsilon(x')\rangle=0 \) and \( \langle \epsilon(x)\epsilon^\ast(x')\rangle=2w^2\delta_g(x,x') \) with \( \delta_g(x,x') \) being the covariant delta function in the Poincar\'{e} disk and \( w \) measuring the degree of randomness.
\par

The continuum action~\eqref{eq_7} can be discretized on regular \( \{p,q\}\) tilings of the unit disk, i.e., tessellations by regular \(p\)-gons with \(q\) such polygons meeting at each vertex, subject to the hyperbolicity condition Eq.~\eqref{eq_1}. On such a \( \{p,q\}\) tiling lattice, the discretized action reads
\begin{equation}
    \begin{gathered}
        S[\{\psi_\mu\}] \simeq \frac{qA}{2p} \sum_{\mu,\nu\in \mathcal V_L} \psi^\ast_{\mu}\bigl( -\mathcal{L}_{\mu\nu} + V_{\mu\nu} \bigr)  \psi_\nu
    \end{gathered}\label{eq_8}
\end{equation}
where \( \mathcal{L}_{\mu\nu} \) is the graph Laplacian of the underlying lattice \( \mathcal{G}_L \), defined by \( \mathcal{L}_{\mu\nu}=d_{\mu}\delta_{\mu\nu}-H_{0,\mu\nu} \) with \( H_{0,\mu\nu} \) being elements of the adjacency matrix \( \hat{H}_0 \) and \( d_\mu \) being the degree of vertex \( \mu \). \( A \) is the area of a single \(p\)-gon, fixed by the Gauss-Bonnet theorem as \( A=(p-2)\pi-2p\pi/q \) for unit negative curvature. The potential term is \( V_{\mu\nu}=qh^2m^2_\mu\delta_{\mu\nu} \), where the dimensionless constant \( h \) represents the geometric factors arising from the lattice discretization and \( m^2_\mu \) encodes the randomness of on-site potentials. The discrete action~\eqref{eq_8} can be mapped to our lattice Hamiltonian~\eqref{eq_2} through an appropriate rescaling of parameters depending on \( \{p,q\} \)~\cite{supp}.
\par

We then estimate the role of critical fluctuations of the lattice models~\eqref{eq_2} by using a phenomenological coarse-grained Landau-Ginzburg description based on the continuous action~\eqref{eq_7}~\cite{Benedetti_jsm_2015,zinn2021quantum}. For the disorder average, we consider a doubled replicated formulation, in which the two sectors couple separately to \( \epsilon(x) \) and \(\epsilon^\ast(x) \), to retain the non-holomorphic covariance, i.e., \( \langle \epsilon(x)\epsilon^\ast(x')\rangle=2w^2\delta_g(x,x') \). In the long-wavelength approximation, its effect is absorbed into the effective parameters of the coarse-grained scalar-field theory~\cite{supp}. Within this phenomenological description, the Ginzburg ratio scales with the distance \( |\tau| \) to criticality as~\cite{supp}
\begin{equation}
    \text{GI}\sim e^{-\ln{\eta}/(2\sqrt{|\tau|})}/|\tau|,\quad\eta>1,
    \label{eq_9}
\end{equation}
up to a non-universal factor. Since \( \lim_{\tau\to 0}\text{GI}=0\), critical fluctuations are asymptotically suppressed, supporting a self-consistent mean-field description~\cite{zinn2021quantum}. By contrast, we obtain \( {\rm GI}\sim |\tau|^{(d-4)/2} \) in Euclidean space so that the mean-field description is stable only for \(d>4\)~\cite{supp}, which is consistent with the numerical observations for non-Hermitian Anderson transitions~\cite{cWang_prb_2025}.
\par

Recall that Anderson transitions are governed by quantum interference and multifractal wave-function fluctuations~\cite{fEvers_rmp_2008}. Their critical behavior generally requires the treatments like nonlinear sigma models, supersymmetric methods, or replica field theories~\cite{dfriedan_prl_1980,shimami_prp_1980,mZirnbauer_prb_1986,mZirnbauer_nulphy_1986,tRizzo_prb_2024,aaltland_arxiv_2026}. Nevertheless, our scalar-field analysis captures a key geometric mechanism: the exponential volume growth of hyperbolic space suppresses critical fluctuations~\cite{rKrcmar_jpa_2008,zWu_pre_2010,npBreuckmann_pre_2020}. This geometric suppression offers a possible explanation for the mean-field-like critical behavior in Table~\ref{table_1} observed numerically on hyperbolic-like lattices.
\par

\emph{Remarks.}$-$(i)~As the considered non-Hermitian Anderson transition falls within the mean-field regime, it is instructive to compare our results with non-Hermitian Anderson models in Euclidean spaces above the upper critical dimension for the mean-field results (\( d=5,6 > 4\))~\cite{cWang_prb_2025} or on the Bethe lattice (\( d=\infty \))~\cite{jShang_prb_2026}. Consistent with our expectations for mean-field universality, these systems also exhibit \( \nu=1 \). 
\par

(ii)~Hermitian Anderson transitions on hyperbolic-like lattices also exhibit compatible critical exponents, e.g., \(\nu\simeq 1\) for the \(\{8,8\}\) and \(\{8,3\}\) tilings~\cite{aChen_prl_2024}, and \(\nu=1.02\) for the \(\{3,8\}\) and \(\{4,8\}\) tilings~\cite{Li_cp_2024}. The agreement with our non-Hermitian results suggests that negative curvature, rather than Hermiticity or microscopic tiling details, controls the universality, and the same mean-field-dominated critical behavior applies to Hermitian hyperbolic-like lattices. Notably, this sharply contrasts with Euclidean systems~\cite{xLuo_prresearch_2022,cWang_prb_2022,cWang_prb_2023}, where non-Hermiticity can change universality via the extension from the 10-fold to the 38-fold symmetry classification~\cite{aAltland_prb_1997,zGong_prx_2018,kKawabata_prx_2019}.
\par

\emph{Experimental relevance.}$-$The proposed geometric universality is of direct experimental relevance, since negatively curved geometries can be engineered in various artificial platforms, like photonic waveguide arrays~\cite{lHuang_nature_2024} and circuit-QED networks~\cite{kAlicia_nature_2019,iBoettcher_pra_2020,bPrzemyslaw_prl_2022,wZhang_nc_2022}, where disorder is naturally present or deliberately introduced. Our model~\eqref{eq_2} can also be implemented in engineered classical networks, particularly electrical circuits~\cite{pLenggenhager_nc_2022,aChen_nc_2023,wZhang_nc_2023}, where complex on-site disorder can be introduced through random variations in capacitive and inductive impedances. A circuit-based proposal for detecting the geometric universality is provided in Supplemental Material~\cite{supp}.
\par

\emph{Conclusion.}$-$In conclusion, we numerically show that Anderson transitions across different \(\{p,q\}\) tilings on hyperbolic-like lattices display universal critical behavior characterized by a common critical exponent. This observation differs from conventional Euclidean classification governed by symmetry and dimensionality. Instead, it is rooted in the macroscopic negative curvature of the underlying geometry, suggesting the emergence of a geometric universality class.
\par

\begin{acknowledgments}

This work is supported by the National Natural Science Foundation of China under Grant No.~12574023. We also thank the Fundamental and Interdisciplinary Disciplines Breakthrough Plan of the Ministry of Education of China (JYB2025XDXM410), the Natural Science Foundation of China under Grant No.~12375051, and Tianjin University Self-Innovation Fund Extreme Basic Research Project Grant No. 2025XJ22-0014 and No. 2025XJ21-0007.\par

\emph{Data availability.}$-$The raw data supporting the findings of this manuscript are available on Zendo. The custom code for data calculations is available from the authors upon reasonable request.\par

\end{acknowledgments}

\bibliographystyle{apsrev4-2} 
\bibliography{refs}

\end{document}